# Shuffling mode competition leads to directionally-anisotropic mobility of faceted Σ11 boundaries in face centered cubic metals

Megan J. McCarthy [a], Timothy J. Rupert [a,b,*]
[a] Materials and Manufacturing Technology, University of California, Irvine, CA 92697, USA
[b] Department of Materials Science and Engineering, University of California, Irvine, CA 92697, USA
* Email: trupert@uci.edu

Faceted grain boundaries can migrate in interesting and unexpected ways. For example, faceted Σ11 <110> tilt grain boundaries were observed to exhibit mobility values that could be strongly dependent on the direction of migration. In order to understand whether this directionally-anisotropic mobility is a general phenomenon and to isolate mechanistic explanations for this behavior, molecular dynamics simulations of bicrystals evolved under an artificial driving force are used to study interface migration for a range of boundary plane inclination angles and temperatures in multiple face centered cubic metals (Al, Ni, and Cu). We find that directionally-anisotropic mobility is active in a large fraction of these boundaries in Ni and Cu and should therefore impact the coarsening of polycrystalline materials. On the other hand, no such anisotropy is observed in any of the Al boundaries, showing that this behavior is material-dependent. Migration of the faceted boundaries is accomplished through transformation events at facet nodes and incommensurate boundary plane facets, which are termed shuffling modes. Three major shuffling modes have been identified, namely Shockley shuffling, slip plane shuffling, and disordered shuffling. A shift from the first two ordered modes to the third disordered mode is found to be responsible for reducing or removing directionally-anisotropic mobility, especially at the highest temperatures studied.

## A. Introduction

Faceted grain boundaries have unique features that can influence microstructure evolution. The formation and presence of faceted boundaries has been shown to slow the migration of neighboring interfaces [1, 2] while their disappearance from a microstructure in the form of a defaceting transition has been connected to abnormal grain growth [3]. Each sub-structure of a faceted boundary can also have its own unique impact on microstructure. The difference in boundary orientation means that each plane can have very different character and migration behavior in itself [4-6], and the change of boundary plane angle also requires new defects to be formed at the sites adjoining each plane, called facet junctions. Studies of facet junctions have shown them to be unique defects [7], with local strain profiles that can make them preferred segregation sites for impurities and dopants [8, 9]. Understanding the dynamic interplay between the individual components of faceted boundaries as well as their collective action can lead to improved quantitative models of grain boundary network evolution.

Recent molecular dynamics studies of faceted boundaries have shown that a detailed analysis of their motion can provide insight into more general boundary migration behavior. In a study of a faceted mixed-character Σ7 boundary in Al, Hadian et al. [10] undertook an analysis of how each sub-structure (low energy plane, high energy plane, and junction defects) contributed to the overall boundary migration. They found that the primary mechanism, step flow, could be broken down into three different stages, each of which had different velocities and responses to changes in driving force. Their model for understanding dynamic step flow in this faceted boundary has application in understanding the migration of kink motion in general grain boundaries. An additional example of the importance of studying the atomistic mechanisms of faceted boundary motion can be found in a work by Humberson et al. [4]. These authors compared three faceted Σ3 boundaries in Ni, two of which were highly mobile and exhibited anti-thermal mobility (where boundaries migrate faster at lower temperatures than higher ones). While all three boundaries had very similar structures and migrated via movement of the same triplets of Shockley partial dislocations, the kinetics of their motion was heavily influenced by the precise boundary plane inclination of the mobile, higher-energy facet. Specifically, it was found that the orientation of the Shockley dislocations had an unexpectedly higher energy barrier to motion in the thermally-activated boundary as compared to the two highly mobile, anti-thermal interfaces. This finding underscores how understanding faceted boundary migration can highlight subtle but highly impactful differences in the dynamic behavior of interfaces.

To add to the field's growing knowledge of faceted interfaces, we have chosen to focus on Σ11 <110> tilt boundaries, which have long been known to have special energetic and structural features. The Σ11 symmetric boundary on this tilt axis has an unusually low boundary energy [11, 12], second only to the lowest-energy symmetric Σ3 or coherent twin boundary. This low energy has been shown to be related to the Σ11's highly symmetric and compact repeating unit, termed a C unit in the well-known structural unit model [13-15]. The compactness and low energy of the symmetric Σ11 plane are what cause asymmetric Σ11 boundaries to facet readily at boundary plane inclination angles below approximately 60.5°, giving rise to a number of interesting facet plane and facet junction structures [16-18]. Amongst them are junction defects seen in low stacking fault energy materials such as Cu and Au, in which Σ11 boundaries frequently relax via emission of stacking faults from the grain boundary [19-21]. These defects, which extend outside of the boundary plane, have been shown to directly contribute to boundary properties such as increased grain boundary sliding resistance [22] and directionally-dependent mobilities [23]. The latter is particularly interesting, as boundary mobility is typically treated as a constant that does not change with migration direction, but our prior work in Ref. [23] demonstrated that large variations in mobility can be found when a faceted Σ11 boundary moves in opposite directions. However, this initial report of directionally-anisotropic mobility was for a single boundary character in Cu, leaving open questions about the dependence of this mechanism on the nanoscale details of grain boundary structure and uncertainty about whether such behavior was widespread. Answering these questions and exploring the dynamic behavior of faceted Σ11 boundaries presents a unique opportunity to connect nanoscale morphology to novel boundary properties and behavior.

In this work, we perform a broad survey of mobility trends in two dozen faceted Σ11 <110> bicrystals, spanning multiple boundary inclination angles and FCC metals with simulations at three temperatures. We establish that directionally-anisotropic mobility is common to most of the Cu and Ni boundaries studied but only very weakly present in the Al boundaries, as a whole suggesting a new boundary migration trend that must be considered. We find that directional-anisotropy in the mobility of the Cu and Ni boundaries can be connected to the operation of an extra motion mechanism, slip plane shuffling. Differences in stacking fault energy create the structural conditions necessary for directionally-anisotropic mobility, which can explain its very low presence in Al boundaries. Using a cluster tracking algorithm, we then establish that competition between the ordered and disordered migration mechanisms provide a starting point for understanding the temperature- and direction-dependent trends in boundary mobility.

## B. Methods

Visualization of boundaries and parts of the data analysis were performed using the OVITO software toolset [24]. All boundary snapshots are quenched to remove thermal noise and then colored according to OVITO's Adaptive Common Neighbor Analysis algorithm [25], where green indicates local FCC orientation, red indicates HCP orientation, and grey indicating an "unknown" orientation. The lowest energy Σ11 <110> tilt grain boundary is shown in Figure 1(a), with an orientation of $(113)_1/(11\text{-}3)_2$ (note that the viewing angle of all boundary images in this work is down the <110> axis, unless otherwise indicated). From this reference point, we choose four different boundary inclination angles that are spaced approximately 10° apart from each other (boundary inclination angle, $\beta$ = 15.8°, 25.5°, 35.3°, and 46.7°). An example of a Ni-1, $\beta$ = 15.8° boundary at a homologous temperature of $T_H$ = 0.8 is shown with its potential energy profile in Figure 1(b). Additionally, one asymmetric, but not faceted, Σ5 <100> tilt bicrystal was created and tested as a control sample. Three different face centered cubic elements (Al, Cu, and Ni) were selected and each represented with two different embedded-atom model (EAM) potentials, for a total of 6 potentials or "materials" probed in this study. We refer to these potentials as Al-1 [26], Al-2 [27], Cu-1 [28], Cu-2 [29], Ni-1 [30], and Ni-2 [26] in the subsequent text. The variety of metals was selected to sample materials with a shared face centered cubic crystallography but provide variation in other properties such as lattice constant, grain boundary energy, and stacking fault energy. A selection of their properties is included in Table I. Melting temperatures were calculated using the method outlined by Wang et al. [31], and the symmetric Σ11 ($\beta$ = 0°) boundary energies calculated using the methodology of Tschopp et al. [12] outlined in the next paragraph.

Fully periodic simulation cells were generated in the Large-scale Atomic/Molecular Massively Parallel Simulator (LAMMPS) [32] using the boundary minimization code developed by Tschopp et al. [12]. This algorithm probes all possible fully periodic structures for a given crystal orientation by iteratively shifting the boundary plane and deleting different combinations of atoms. The resulting bicrystals with the lowest grain boundary energies for a given inclination angle were selected and used for subsequent simulations. Bicrystal mobility studies in periodic boundaries, and especially faceted boundaries, have been shown to be highly sensitive to system size [4, 5, 33-35]. If the length of the bicrystal in the direction normal to the grain boundary plane ($L_y$ in this study) is not sufficiently long, mobility results can be artificially altered due to interactions across the periodic boundary [33-35]. However, large system sizes become computationally expensive, especially when probing a larger parameter space. We thus endeavored to create bicrystals that were large enough to avoid common mobility artefacts, but small enough to optimize computational efficiency. Using the criteria provided by Deng and Deng [35], we determined that a minimum $L_y$ length of 30.0 nm would be acceptable for the four $\Sigma11$ bicrystals under study (actual range used was 31.4 nm to 42.8 nm). The length of the dimension parallel to the boundary plane, $L_x$, determines the number of repeating units in faceted boundaries (also called the facet periodicity). The four $\Sigma11$ boundaries studied here have facet periodicities that vary significantly. For the minimum periodicity, which yields long repeating structures, 4 repeating units were used, and for the maximum, 12 repeating units were used. This yielded a range of $L_x$ values between 16.7 nm and 25.0 nm. The minimum tilt-axis thickness, $L_z$, was fixed to be 7 repeats of the lattice parameter $a_0$, giving a range of 3.6 nm to 4.1 nm across the potentials. These size parameters resulted in 24 bicrystals with each containing ~150,000 to ~250,000 atoms.

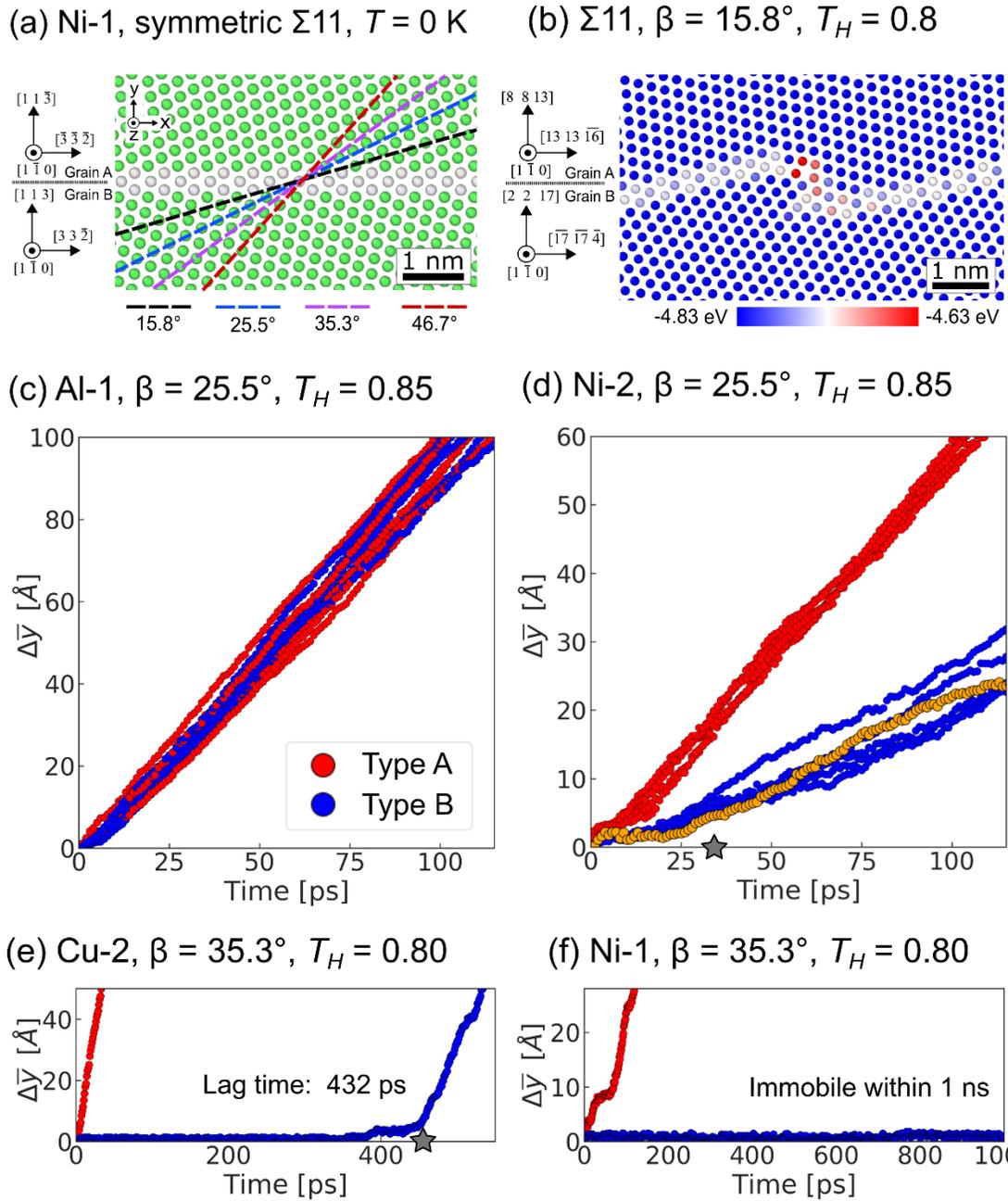

**Figure 1.** (a) Symmetric Σ11 <110> tilt boundary. Asymmetric boundaries were created by varying the boundary plane inclination angle β from the 0° symmetric configuration to the four angles indicated by the colored dashed lines. (b) Potential energy profile of a faceted boundary with β = 15.8° (black dashed line in (a)). (c) Average boundary displacement versus time in which Type A (red) and Type B (blue) have the same mobilities (i.e. slopes are the same). (d-f) Examples of directionally-anisotropic mobility, where the red and blue slopes are different. Many Type B boundaries have significant immobile phases ($t_{lag}$), as indicated by grey stars. An example of a simulation run with a lag time in (d) is highlighted in orange. (f) A Ni-1 boundary that is immobile within the time frame chosen for this work (1 ns).

| Potential Name | $a_0$ [Å]* | $\gamma_{SF}$ [mJ/m$^2$]* | $T_{melt}$ [K] [31] | Energy, $\Sigma11$, $\beta = 0°$ [mJ/m$^2$] |
|---|---|---|---|---|
| Cu-1 [29] | 3.639 | 26 | 1352 | 353.3 |
| Cu-2 [28] | 3.615 | 44 | 1357 | 309.9 |
| Ni-1 [30] | 3.52 | 82 | 1995 | 533.1 |
| Ni-2 [26] | 3.52 | 103 | 1635 | 465.9 |
| Al-1 [26] | 4.05 | 117 | 850 | 104.1 |
| Al-2 [27] | 4.05 | 146 | 1035 | 150.8 |
| * Values taken from reference in first column | | | | |

**Table I.** Basic properties of the six EAM potentials utilized in this study. Lattice constants and stacking fault energies were taken from the references for each potential, included in the first column. Melting temperatures and boundary energies were calculated using the methods of Wang et al. [31] and Tschopp et al. [12], respectively.

All simulations were run using an NPT ensemble at three different $T_H$ values of 0.8, 0.85, and 0.9. Temperature was regulated and a zero pressure was maintained using a Nosé-Hoover thermostat and barostat. Annealing runs were first initiated at half the target temperature by giving each atom a randomized velocity. The system was then allowed to ramp over approximately 20 ps to the target temperature, then held at this temperature for a further 100 ps to allow for relaxation of the interfacial structure. At least 6 unique (in terms of initial randomized velocity) configurations were run for each temperature-potential-bicrystal combination.

After relaxation, the energy-conserving orientational artificial driving force (ADF) developed by Ulomek et al. [36] was used to drive boundary motion. To capture first and second nearest neighbors, a cutoff value of 1.1 $a_0$ was selected. The ADF was applied for a minimum of 120 ps and a maximum of 1 ns. ADFs induce boundary motion by introducing an energy gradient across the bicrystal, through the assignment of potential energies to each atom based on their local orientation. This is analogous to experimental techniques employing a magnetic driving force [37, 38], as one grain is favored to grow at the expense of the other. Bulk atoms in the favored grain are given potential energy values of $-\Delta E/2$, and atoms in the bulk of the other, unfavored grain are given values of $\Delta E/2$, resulting in a potential energy difference with an average magnitude of $\Delta E$ (in units of eV/atom). Grain boundary atoms, which are assigned energy values between those of the two neighboring grains, are driven to shift towards the lower-energy orientation. Over time, these shifts allow the favored grain to grow at the expense of the other. In this study, we probe the migration of each boundary in two opposite directions, which is accomplished by swapping the selection of the favored grain in the ADF framework. The results show that this swapping results in very different migration behaviors in the majority of cases, meaning that we need different terminology for migration in each direction. We thus refer to migration/growth of Grain A as a Type A migration. To ensure clarity, the visualization of boundaries will only show the topmost boundary of each bicrystal, meaning that Grain A will always appear above Grain B, although we note that all actual measurements were made using both boundaries. Type A migration will thus always imply that the boundary shown in any figure is moving in the negative Y-direction. The opposite is true for Type B migration, which will always refer to the bottom grain in snapshots growing, meaning that the boundary moves in the positive Y-direction.

Similar to problems with system size in molecular dynamics, periodic boundaries using ADFs can be sensitive to $\Delta E$, the driving force energy value [5, 33-35]. Race et. al [39] have demonstrated that high driving forces do not affect the energy barriers of atomistic migration mechanisms in defect-heavy boundaries. Based on the stated upper limit by these authors, $\Delta E$ values between 10-25 meV/atom were tested initially. The highest value of 25 meV/atom was chosen because a number of boundaries proved to be very slow. This is consistent molecular dynamics studies of faceted and non-faceted Σ3 boundaries on the <110> tilt axis [4, 40], which also required higher driving forces to induce appreciable motion due to lattice friction effects.

Grain boundary velocities were measured by tracking the mean boundary position of each of the two boundaries separately and calculating the average displacement from their starting position. All simulations were run for at least 120 ps, which was adequate to obtain a steady-state velocity for the majority of bicrystals. Tracking was stopped either when the favored (growing) grain had consumed approximately 75% of the unfavored grain (to avoid boundary interactions) or when 1 ns had been reached, whichever came first. However, many boundaries in this data set are essentially immobile or begin moving only after a long lag time. To ensure a steady-state velocity for boundaries with significant lag times, collection of boundary statistics was begun only after the averaged boundary position had changed by an amount greater than twice the highest displacement recorded in the immobile phase, leading to values between 5–10 Å. Statistics for the boundaries that were relatively immobile were sampled from the final 100 ps of the simulation. Mobility was then calculated using

$$M = \frac{v}{P} \tag{1}$$

where $v$ represents the boundary's velocity, $P$ is the driving pressure experienced by the boundary (in this case, through the ADF), and $M$ is the term that relates the two, with units of m/(GPa·s). $P$ is related to the ADF's energy gradient value, $\Delta E$, through the average atomic volume, $\Omega$ (found using the lattice constants listed for each potential in Table I), and the conversion of eV to GPa, such that

$$P = \frac{\Delta E}{\Omega} \cdot \frac{1.602 \cdot 10^{-28} \, GPa}{eV}. \tag{2}$$

**C. Results and Discussion**

*1. Trends in directionally-anisotropic mobility*

Figure 1(c-f) shows the mean Y-direction displacement, $\Delta \bar{y}$, plotted against time to show the different migration patterns that are observed in this study. The top two plots feature boundary trajectories of Al-1 (Figure 1(c)) and Ni-2 (Figure 1(d)) at β = 25.5° and $T_H$ = 0.85. The red and blue curves show the displacements for Type A and Type B migration, respectively. Though the Al-1 and Ni-2 faceted boundaries have macroscopically identical crystallography and the same simulation parameters, there are significant differences in their migration behaviors. For the Al-1 data in Figure 1(c), the slopes of the red and blue curves are very similar to each other, indicating that Type A and Type B migration have similar velocities, and thus mobilities since the same driving force was used in each direction. By contrast, the slopes of the red and blue curves in the Σ11 Ni-2 data in Figure 1(d) are very different from each other. In this case, Type A migration proceeds at a much faster rate (2-3.3 times faster) than Type B migration.

In addition, there are differences between the materials in how migration begins. In the Al-1 example (Figure 1(c)), all boundaries moved a significant distance within a few picoseconds of application of the ADF. However, in the Ni-2 example (Figure 1(d)), there is a considerable delay between ADF application and the onset of motion, which can be quantified as a lag time, $t_{lag}$. It is especially apparent in the blue Type B trajectory lines but can also be observed to a more modest extent in the red Type A data as well. The simulation run with the longest lag time of approximately 35 ps (gray star) has been highlighted in orange after the boundary position has changed by approximately 5 Å. Figure 1(e) and (f) show two other common trends in lag time observed throughout the Cu and Ni dataset. Figure 1(e) shows a Cu-2 boundary at $\beta = 35.3°$ and $T_H = 0.8$ with a lag time of 432 ps, which was typical for some inclination angles and lower temperatures. Several other boundaries appear to be either completely immobile or have lag times beyond the 1 ns run time limit, such as the example trajectory for Ni-1 at $\beta = 35.3°$ in Figure 1(f).

The existence of boundary migration that is different in the two opposite directions means that two values mobility values, $M_A$ and $M_B$, are needed based on the direction of boundary migration. To measure the magnitude of difference between Type A and Type B mobilities, we define the anisotropy ratio A:

$$A = max\left(\frac{M_A}{M_B}, \frac{M_B}{M_A}\right) \qquad (3)$$

In this form, $A$ expresses how many times faster motion in one direction is than in the other and would therefore be close to 1 for boundaries with similar mobilities, such as Al-1 in Figure 1(c) where $A = 1.05$. In contrast, the Ni-2 boundary of Figure 1(d) has a significantly lower Type B mobility, which results in a larger value for the anisotropy ratio of 2.21.

Figure 2 presents the mobility data for all materials, temperatures, and inclination angles used in this study, presented in three different columns. Each row represents one potential, with the stacking fault energy indicated underneath the potential label. Within each plot, the X-axis displays the inclination angle $\beta$ in degrees and the three data sets in each plot corresponding to the three homologous temperatures that were tested ($T_H$ = 0.8, 0.85, 0.9). The Y-axes of the left-most and center columns show the calculated mobility values for Type A and Type B migration ($M_A$ and $M_B$), respectively. The right-most column contains the anisotropy ratio $A$ from Equation 3, shown on a log scale. Error bars here and in all future plots show the standard deviation around the mean value. Figure 2 reveals that the anisotropic mobility seen in the Ni-2 boundary in Figure 1(d) is observed for many of the Cu and Ni boundaries. In fact, for Cu and Ni isotropic mobility is the exception, only observed in Cu-1 and Cu-2 at $\beta$ = 15.8° for all three temperatures and at $\beta$ = 46.7° for the highest temperature. In the majority of other angle-potential combinations of Cu and Ni, Type B mobility is considerably lower than Type A mobility. In the case of Ni-1 in Figure 2(c), at $T_H$ = 0.8 there are even multiple examples of very sluggish boundaries at $\beta$ = 25.5° and 35.3°, leading to extreme anisotropy values of ~202 and ~50, respectively, which are outside the limits of the plot. In contrast, the two Al potentials have far lower anisotropy values overall, being close to 1 in most cases and never higher than 1.36.

    Though a thorough investigation of the effect of temperature is outside the scope of this work, some general mobility trends and thus also anisotropy trends can be established with the three homologous temperatures studied. Increased temperature generally increases the mobility of both Type A and Type B-driven boundaries, suggesting that both migrate via thermally-activated mechanisms [41]. However, mobility increases more quickly in the Type B boundaries, meaning that anisotropy values generally decrease with increasing temperature, though commonly

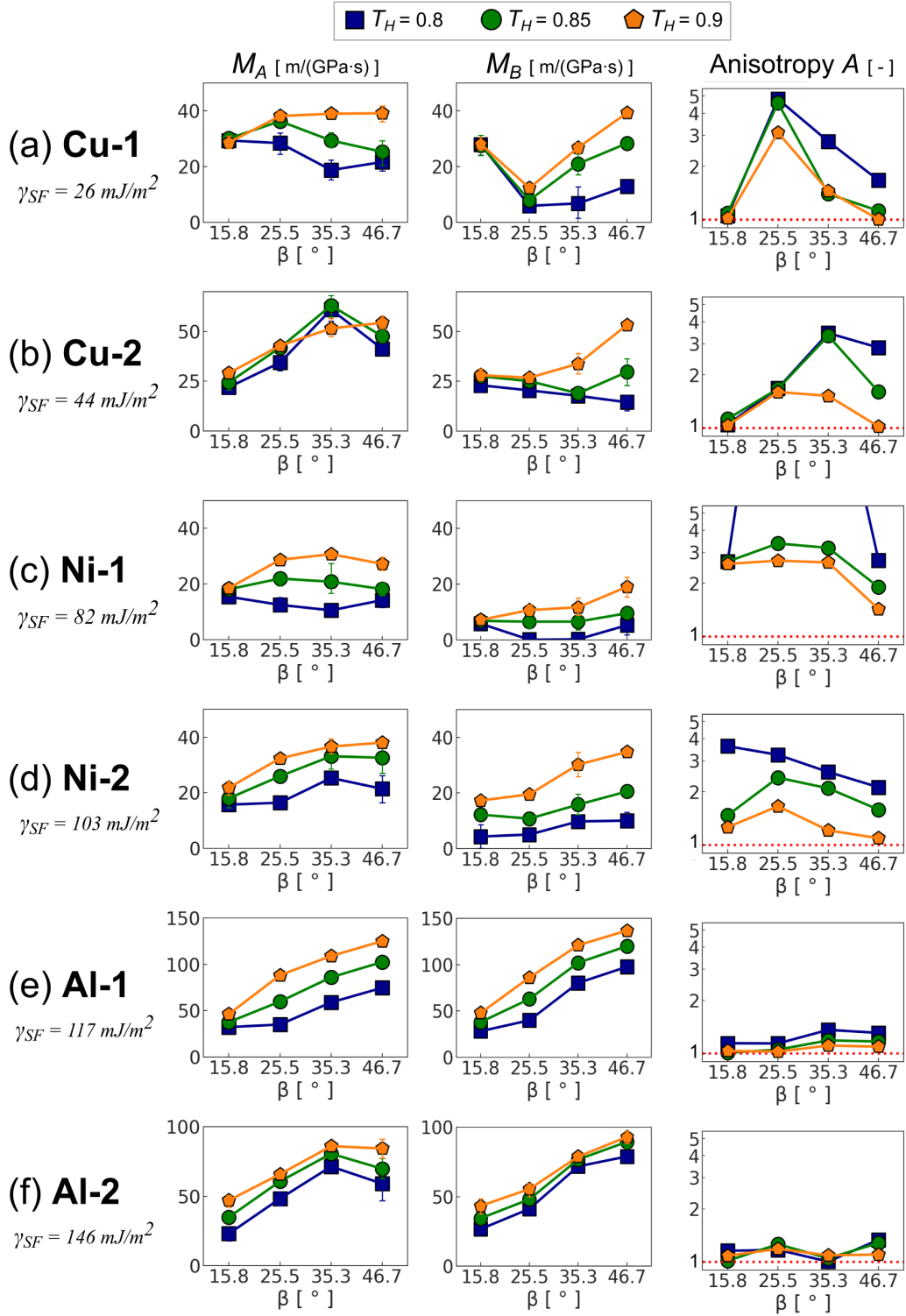

**Figure 2.** Type A mobility (leftmost column), Type B mobility (center column), and anisotropy ratios (rightmost column) for each potential (rows), plotted against the inclination angle β (X-axis) and showing the data for three different homologous temperatures (different-colored curves).

remain elevated even at the highest homologous temperature of 0.9 (orange curves). The most obvious exceptions to this trend are the β = 15.8° boundaries, where mobility is insensitive to temperature for both Cu potentials and the Ni-1 potential. In addition, a unique example of thermal dampening is seen in Cu-2 at β = 35.3° (Figure 2(b), first column), where higher temperatures lead to lower boundary mobility for Type A motion.

As mentioned earlier and shown in Figure 1(b-f), almost all of the boundaries with elevated anisotropy values also had a significant phase of immobility in Type B migration before reaching a steady-state velocity. Figure 3 presents these Type B lag times as a function of inclination angle for the Cu and Ni boundaries, with the colors again indicating the different homologous temperatures. The lag times track fairly well qualitatively with the trends in anisotropy for each homologous temperature. The angles and temperatures with peak anisotropy values also tend to be those with the longest or near-longest lag times measured. Also of interest are temperatures and angles where, despite having low anisotropy values, the lag times are still elevated, for example at 46.7° in Cu-1. Relationships between lag times and boundary structure will be discussed in Section C.3.

As mentioned in the Methods section, an identical set of simulations were run on an asymmetric Σ5 tilt bicrystal in order to provide a control data set. Both the Σ5 and Σ11 boundaries are asymmetric, but the Σ5 lacks the ordered nature of facet planes and facet junctions. The results of these auxiliary simulations are not shown but gave anisotropy ranges between 1.00 and 1.09 for Σ5 boundaries over all homologous temperatures and across all potentials studied. Based on those results, a value of $A = 1.1$ (or a difference of 10% in mobility) was chosen to represent a boundary between directionally-anisotropic and isotropic mobility in the faceted Σ11 boundaries. By this standard, it is clear that the great majority of Cu and Ni boundaries shown in Figure 2 can be categorized as strongly anisotropic.

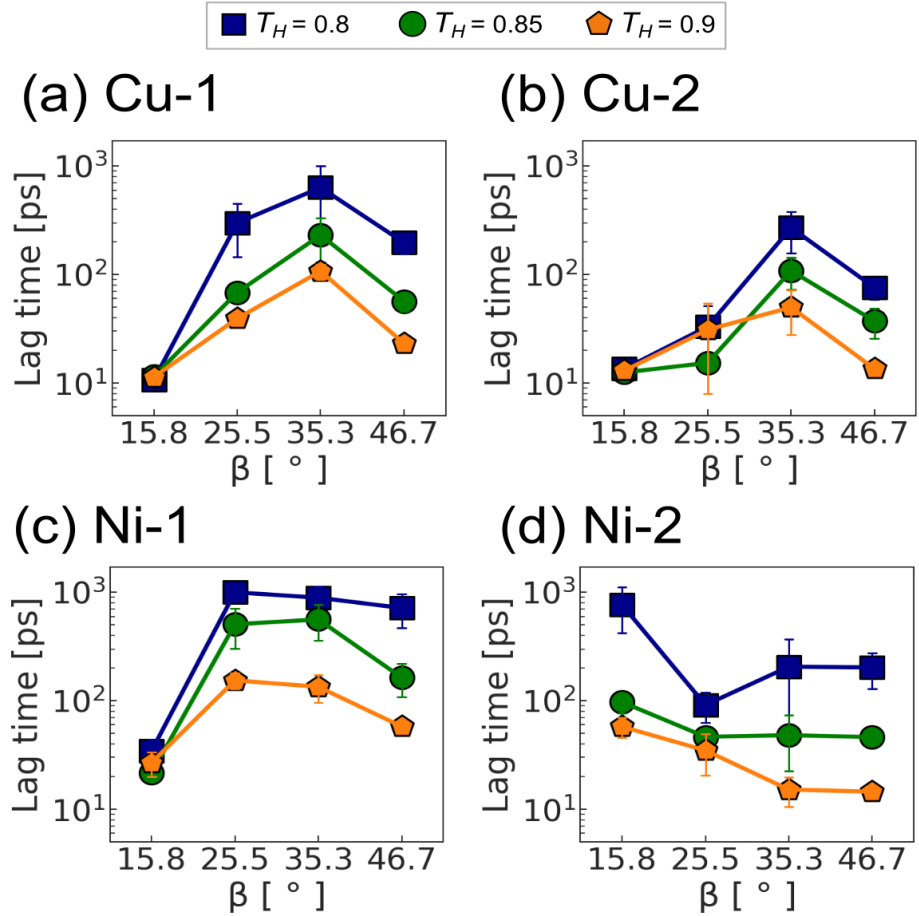

**Figure 3.** Lag times ($t_{lag}$) as a function of inclination angle β for Type B-migrating boundaries in (a, b) Cu and (c, d) Ni boundaries, for three different homologous temperatures.

*2. Common structures of Σ11 boundaries*

The 72 unique Σ11 bicrystals included in this dataset all share a small number of characteristic units that remain present both during annealing and migration, even at the highest homologous temperature of $T_H = 0.9$. These units agree well with structures defined in the structural unit model (SUM) [15], so we will utilize this formalism here as well. Because of the similarities between all boundaries, a small selection are used to introduce the entire set. In several following figures, boundary snapshots will be accompanied by schematics that highlight important details such as important crystallographic planes. To better show these planes, atoms in the schematics will be shaded by approximate plane height relative to the tilt axis, with darker atoms one {110} plane height lower than lighter ones. Figure 4 shows two facet periods of as-annealed ($T_H$ = 0.8) boundaries for two different materials at β = 25.5°. Starting with Al-2 in Figure 4(a), the ascending segment is oriented along a plane which we have already seen in Figure 1(a), the Σ11 symmetric boundary plane (SBP). SBPs consist of a chain of diamond-shaped structural units called C units, which are denoted in the schematic on the right side of Figure 4(a). The descending segment in Figure 4(a), which appears rougher than the SBP facet, also consists of a pair of structural units from the SUM called E units. E units are comprised of 6 atomic columns enclosing a region of excess free volume in their centers, making them the largest structural unit that remains fully ordered [42, 43]. Their kite-like shape can come in several variations but given the challenge of distinguishing very subtle variations in structure at high temperature during migration, we will refer to all of them simply as E units. The schematic on the right side of Figure 4(a) shows the E units outlined in red. It is important to note that the Adaptive Common Neighbor Analysis algorithm in OVITO does not always identify all 6 atomic columns of the E unit as 'other' type (colored white in snapshots). A common occurrence in E units both during annealing and migration is the dissociation of one or more atomic columns, shown in Figure 4(a) at the right-most bracket edge and in isometric view in Figure 4(c). Dissociated E units are indicated in schematics with dashed red lines.

Figure 4(b) shows the crystallographically-identical boundary in a different material, Cu-2. Instead of two relatively simple ascending and descending boundary segments as seen in the Al-2 boundary of Figure 4(a), there are 3 different segments in the Cu-2. The first of them is also an SBP facet, made of three C units instead of the five seen in the Al-2 boundary. The second feature is a terrace-like flat segment, which is a secondary facet of the Σ11 boundaries oriented along $(111)_A/(001)_B$ plane. More details on this plane, which is an incommensurate boundary with special crystallographic properties, can be found in Brown and Mishin's in-depth study of the faceting behavior of Σ11 boundaries [17]. These terrace-like segments are referred to as incommensurate boundary plane (IBP) facets. The third feature which connects the IBP facets to the SBP facets are actually very short stacking faults. These stacking faults are a result of a grain boundary relaxation mechanism called grain boundary stacking fault emission, where a perfect grain boundary dislocation dissociates into two partial dislocations, and is commonly seen in many asymmetric <110> tilt boundaries (including Σ11) [12, 13, 43]. As shown in the isometric view of the boundary in Figure 4(d), OVITO's dislocation analysis algorithm [44] identifies the ends of these defects as Shockley partial dislocations (vertical green line) with Burger's vectors of $\mathbf{b}$ = [-1 2 -1] (blue arrow, scaled up by 2.5).

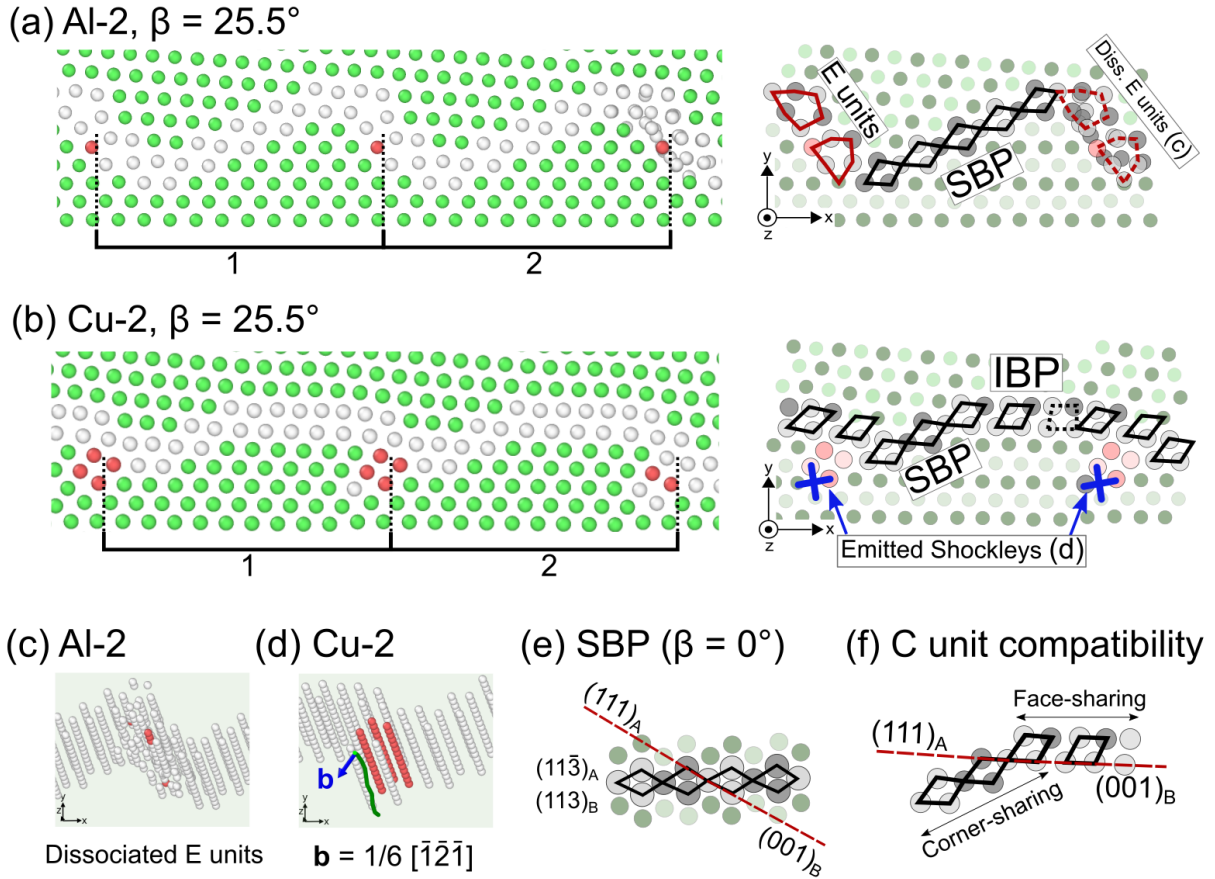

**Figure 4.** Representative structures for (a) an Al-2 boundary and (b) a Cu-2 boundary, each at β = 25.5° and $T_H$ = 0.8. To the right of each are schematics showing the characteristic structural units for each boundary type. (a) The Al boundaries are made of pairs of E units (red outlines) connected by a chain of corner-sharing C units (black diamonds). Atomic columns within E units can dissociate, as shown in the E units on the rightmost side of (a) and shown more in detail in the isometric view of (c). (b) In addition to SBP and IBP segments, Cu and Ni boundaries emit Shockley partial dislocations at facet nodes (blue X). An example of one in isometric view is shown in (d). Cu and Ni boundaries also generally have a higher fraction of IBP facet segment formation, which can also be described using C units in a face-sharing configuration as shown in (f).

Though they appear quite different from one another, the SBP and IBP are closely related to each other structurally, which has important implications for mobility. C units in SBPs are connected together along the corners of their longest axes, as shown in Figure 4(e). If the orientation of C units is changed from corner-sharing to face-sharing (Figure 4(f)), one is left with the same configuration found in the schematic in Figure 4(b). The fact that the C unit is shared between both facet types means that the transitions left-to-right from an SBP facet into an IBP facet are relatively smooth. In contrast, the transition from an IBP facet to an SBP facet is considerably more structurally complex, encompassing the emitted stacking fault. The C unit compatibility, along with the fact that stacking fault emission only occurs into one grain (which is not always the case [22]), gives most Cu and Ni boundaries a distinct geometric directionality, the effects of which will be explored further when examining specific migration mechanisms in Section .3.

Figure 4 demonstrates that crystallographically-identical boundaries in different materials (i.e. Al and Cu) can lead to a variety of different structures. Due to this variety (which is only amplified when boundaries are in motion), the interpretation of a facet plane versus a facet junction or facet junction defect is at times ambiguous. The E unit pairs of the Al boundary could be seen as short facet planes, as steps, or even as two adjacent junctions, depending on the definition. In the case of the Cu boundary, though the smooth transition between SBP and IBP facets is also technically a junction, it plays a less clear role as such than the emitted Shockley partial does. For these reasons, we opt for a slightly different terminology for the major structures in these faceted boundaries, based on the most prominent defect in each boundary type, namely E units and emitted Shockley partials. We will call the mean X-axis position of these defects facet nodes (indicated by the dotted lines above brackets in Figure 4(a) and (b)) and call each defect individually a facet node defect. The term facet segment applies only to IBP and SBP facets.

The plot in Figure 5 illustrates these relationships between anisotropy (Y-axis), material and stacking fault energy (X-axis), and inclination angle (symbols/colors) for $T_H = 0.8$. Potentials are placed in order of increasing stacking fault energy. The two Ni-1 markers which are off the chart are a reminder of the two very high anisotropy values for $\beta = 25.5°$ and $35.3°$ that exist outside the Y-axis limits (at ~202 and ~50, respectively). The highest values of anisotropy exist on the left side of the plot in the Cu and Ni materials, while only very low anisotropy is found for the two Al potentials on the right side (separated by a dotted grey line). This separation clearly corresponds to the main structural difference between the anisotropic boundaries and the reasonably isotropic Al boundaries, namely stacking fault emission at facet nodes. Beyond this critical difference, the relationship between anisotropy and stacking fault energy does not appear to be straightforward. For example, the Ni potentials, which both have higher stacking fault energies than Cu, have consistently higher anisotropy values on average. There also do not appear to be any clear relationships between inclination angle and anisotropy. While only $T_H = 0.8$ is shown in Figure 5, the same conclusions are drawn from the anisotropy data for the other temperatures. This suggests that stacking fault energy may only contribute to anisotropy in allowing phenomena such as grain boundary stacking fault emission to take place. The magnitude of anisotropy likely depends on other geometric or material-dependent parameters that come into play during boundary migration, rather than the static structures of the interfaces by themselves.

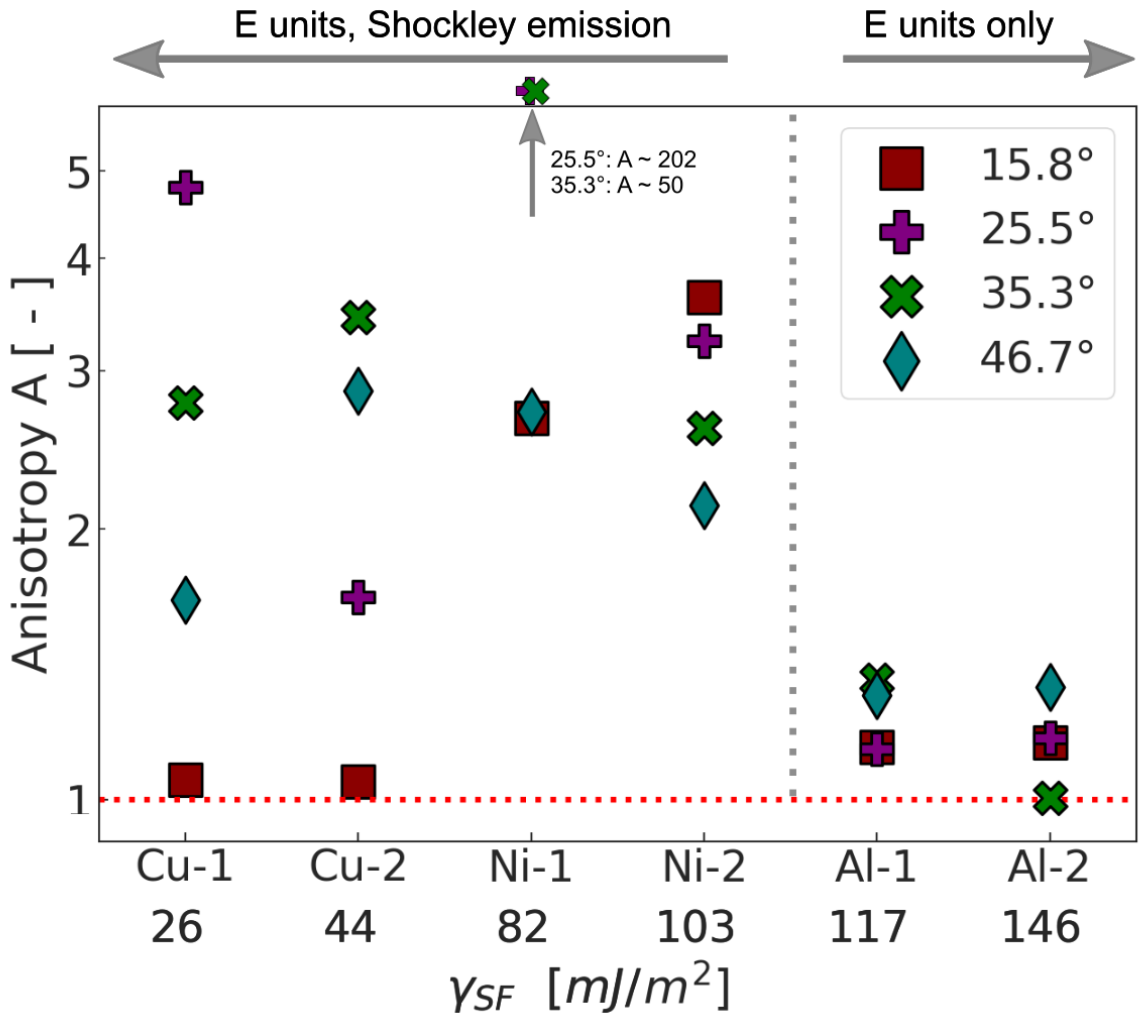

**Figure 5.** Anisotropy ratio values for $T_H = 0.8$ for different materials/potentials, placed in order of increasing stacking fault energy and with the inclination angle shown by the data symbols. The Cu and Ni potentials on the left side have low enough stacking fault energies to be able to emit Shockley partials at facet nodes and have higher overall anisotropies than the two Al potentials, which only have E units at facet nodes. The exceptionally high anisotropy values for $\beta = 25.5°$ and $35.3°$ for Ni-1 lie out of the bounds of the Y-axis.

*3. Shuffling modes, and directionally-anisotropic mobility*

Observation of migrating Σ11 boundaries reveals that it is the motion of facet nodes that provides the most boundary displacement over time. Figure 6(a) shows a snapshot of a Cu-1, β = 35.3° boundary undergoing Type B (slower) motion at $T_H = 0.85$. Facet nodes are indicated once more by dotted lines at the ends of brackets. E units have been outlined as earlier in red lines (with dashed red lines for dissociated E units) and emitted Shockley partials have been marked with a blue X. Unlike the as-annealed facet node structures seen in the Al and Cu boundaries, this snapshot demonstrates that the E unit node, the dissociated E unit (leftmost junction), and the emitted Shockley node can be present simultaneously in migrating boundaries. This implies that structural transformations have occurred in which, for example, an E unit changes into a Shockley partial, or vice versa. We will call the set of boundary transformations which results in one or another type of node forming a *shuffling mode*.

Three primary shuffling modes have been identified, two of which occur only in Cu and Ni due to their lower stacking fault energies. The most common of the two, shown in Figure 6(b), is a cycle involving the emission and contraction of Shockley partials from facet nodes. It begins when the top-most E unit at a node releases a Shockley partial dislocation (Figure 6(b), 5 ps). Returning the emitted Shockley to the node, also called Shockley contraction, leads to the formation of a new pair of E units one (1-1-1)$_B$ plane (the Shockley partial emission plane, parallel to the grey reference line) to the left of the previous E unit. This cycle can be referred to as *Shockley shuffling*. The other shuffling mode unique to Cu and Ni boundaries occurs almost exclusively during Type A motion. An example is shown in Figure 6(c) for a Cu-2 boundary (β = 35.3° at $T_H = 0.8$). If a Shockley partial has been emitted at a node and that node has an IBP facet to its right side as is the case in the second facet node of Figure 6(a), the facet itself can migrate without requiring Shockley partial contraction to occur beforehand. This migration is accomplished by unfolding its C units from a face-sharing configuration into the corner-sharing one of the SBP in the manner shown from 0 ps to 4 ps in Figure 6(c). Essential for this process is the presence of the (111)$_A$ slip plane parallel to the boundary, which is part of why the transition from face-sharing to corner-sharing C unit is relatively easy to accomplish. This motion mechanism can be called *slip plane shuffling*.

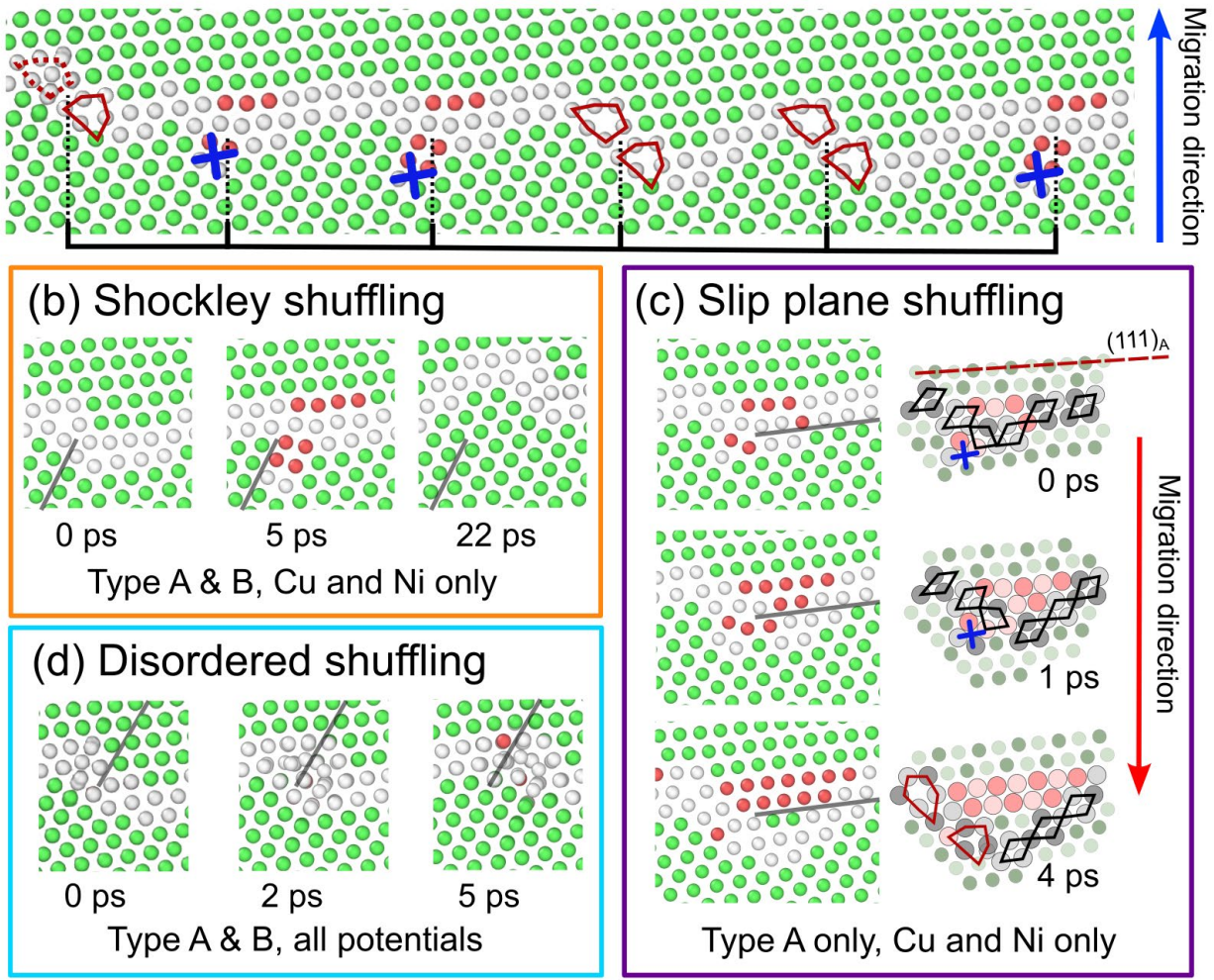

**Figure 6.** (a) Snapshot of a Cu-1 boundary during Type-B migration at $T_H = 0.85$ containing the three different facet node types: E units (red outlines), dissociated E units (dashed red outline on left-most node), and emitted Shockley dislocations (blue X). (b) An example of Shockley shuffling. Note the delay between Shockley emission at 5 ps and Shockley contraction, 17 ps later at 22 ps. (c) An example of slip plane shuffling, in which an IBP facet transforms into an SBP facet. In certain boundaries, a stacking fault (red atoms) in Grain A can also form between two facet nodes. This mode is only possible during Type A motion in Cu and Ni boundaries. (d) An example of disordered shuffling beginning at an E unit after an atomic column dissociation.

The third shuffling mode is what can be called *disordered shuffling*, which begins when one or more columns of an E unit in the facet node dissociates as shown in Figure 6(d). The initial dissociation of one E unit at 0 ps leads to a cascade of dissociations throughout the rest of the facet node over the following few ps, forming a cluster of disordered atoms. Observing the progression of the migrating Cu-1 cluster shown in the snapshots of Figure 6(d) shows how it shifts approximately three (1-1-1)$_A$ planes to the right over the course of 5 ps, with the black bar indicating the original position of the facet node. Once formed, these disordered clusters can exist for varying amounts of time before re-associating into E units. From that point, those E units in Cu and Ni may migrate either via Shockley shuffling (which can in turn result in slip plane shuffling during Type A migration, if circumstances favor it) or dissociate once more and begin a new phase of disordered shuffling. Though all three shuffling modes can be observed in the Cu and Ni boundaries during Type A migration, and disordered and Shockley shuffling are observed during Type B, disordered shuffling is the only mode available to facet nodes in Al boundaries in either direction.

Directionally-anisotropic mobility is a direct result of slip plane shuffling being available only during Type A migration. This is because slip plane shuffling allows the affected node to lower the activation energy for subsequent step migration through Shockley partial contraction. While this is consistent with the findings of Ref. [23] which only reported on a single boundary, the observation of the extended dataset allows us to confirm that these mechanisms are active in the other Cu and Ni boundaries and at multiple inclination angles. Figure 7 shows a facet node transformation taking place in the two different migration directions for an anisotropic Ni-1 β = 15.8° boundary. The three snapshots in Figure 7(a) show the transformation for a node undergoing Type A migration, while the bottom three in Figure 7(b) show it for Type B. The colored arrows between snapshots indicate the direction of migration while the dotted gray lines are fiducial markers denoting the starting positions of the nodes. The initial state at 0 ps of the facet nodes in the first panels of Figures 7(a) and (b) are virtually identical, with IBP facets and emitted Shockley partials. The final states at 2 ps in the third panels are also extremely similar, with both junctions having transformed into a pair of E units. However, looking at the transition states at 1 ps shows that Shockley partial contraction is occurring at different times in each transformation and thus under different conditions. With pure Shockley shuffling shown in 7(b), the contraction takes place in the first half of the transformation. Note that, though only Type B migration is shown here, the same transformation can occur through pure Shockley shuffling (i.e., without slip plane shuffling) during Type A migration as well, but in the reverse order. In contrast, during slip plane shuffling, Shockley contraction takes place in the second half of the transformation, after the process of C unit unfolding has already begun transforming the neighboring IBP facet into an SBP facet.

The significance of this asymmetry in the timing of Shockley partial contraction becomes clear when observing the initial portion of Cu and Ni simulations, where the as-annealed boundaries have primarily emitted Shockley partials and IBP facets are present at almost every facet node. Recall that most Type B-migrating boundaries have extended immobile phases like those shown in Figure 1(d-f) and Figure 3. Evidence from a hybrid experimental and computational study by Bowers et al. [45] on defects very similar to E units in Au strongly suggests that Shockley partial contraction is a thermally-activated process. Thus, all facet nodes in the initial structure of a Cu or Ni boundary will be pinned until the activation energy barrier for contraction is overcome. In contrast, the identical as-annealed boundary can immediately begin structural transformations with a Type A driving force, since it will initiate migration in IBP facets before requiring facet nodes to unpin. The final Shockley contraction is then easier to accomplish because the majority of local structural transformation (the creation of a new segment of SBP facet) has already occurred. To illustrate this process, displacement maps based on the starting and ending positions of each transformation are shown in Figure 7(c) for slip plane shuffling and Figure 7(d) for Shockley shuffling. Displacement vectors are colored by their angle relative to the Y-axis, with red and blue indicating motion in the positive and negative Y-directions, respectively, and green arrows indicating motion along the X-axis. The vectors have also been scaled by a factor of 2.5 to enhance visualization. For slip plane shuffling in Figure 7(d), the largest displacement vectors are those directed along the IBP facet, which is not only already in Grain A but is also oriented along the $(1-1-1)_A$ slip plane. Figure 7(d) shows how this process is quite different in Shockley shuffling. The contraction must both immediately transform a region of Grain A and also acts perpendicular to the same $(1-1-1)_A$ slip plane (red arrows) that enable slip plane shuffling. Though the impact of these differences in timing and displacement are most obvious at the start of a simulation, where most Type A facet nodes move due to slip plane shuffling, the same process is also occurring at different points during Type A migration when IBP facets are present near facet nodes. The cumulative effect of this is manifested in the generally faster migration velocities seen in Type A moving boundaries in Cu and Ni.

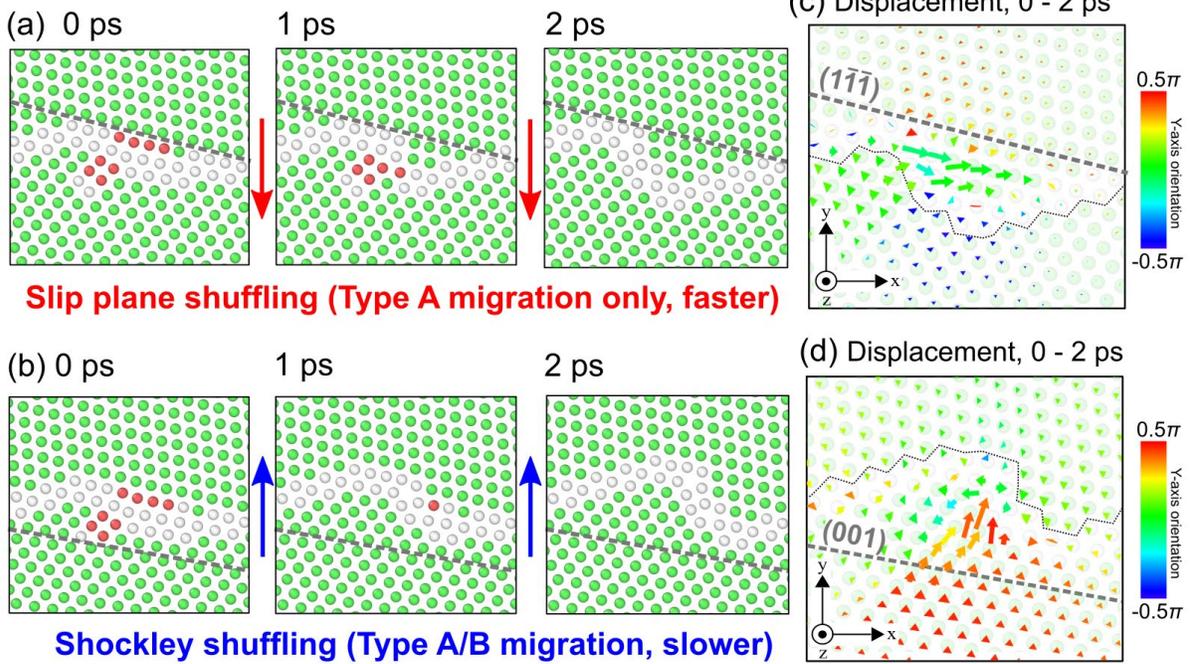

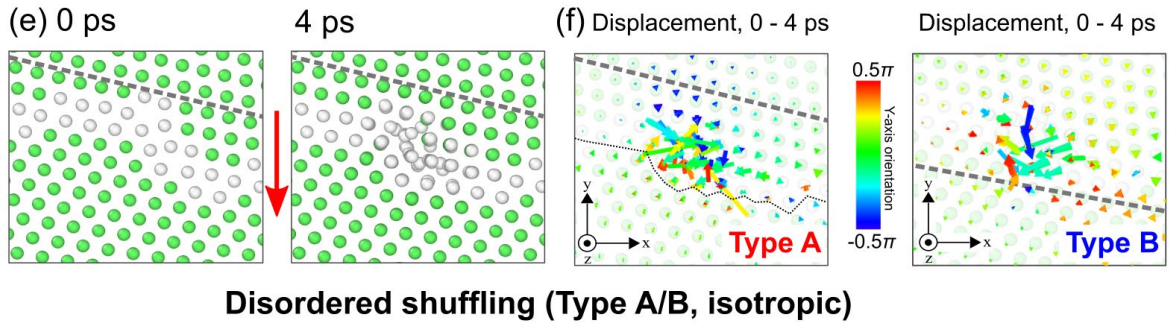

**Figure 7.** (a, b) Comparison of a node transformation that starts and ends in the same form in (a) Type A and (b) Type B migration but have different intermediate stages and (c, d) displacement histories. Shockley partial contraction also occurs at different times in each transformation. (e) Illustration of disordered shuffling during Type A motion. Its displacement map in (e) closely resembles that of a similar 4 ps displacement in Type B. All displacement vectors have been scaled by 2.5 for visualization and are colored by their orientation relative to the Y-axis.

Comparing the displacement maps of Shockley and slip plane shuffling to those illustrating disordered shuffling offers an explanation as to how directionally-anisotropic mobility is avoided in Al. Figure 7(e) shows a facet node moving via disordered shuffling in the Type A direction (red arrow between images) over the course of 4 ps, with the associated displacement map shown in Figure 7(f). The second displacement map in Figure 7(f) was created using a very similar disordered shuffling event occurring over 4 ps, but in the Type B direction. The two disordered shuffling displacement maps in this figure are notably different to those showing slip plane and Shockley shuffling. Atomic column dissociation means that individual atoms can move and hop relatively easily within the free volume of E units, leading to crossing of displacement vectors in the map and a clustered appearance, demonstrating that shuffling is highly localized to the area immediately within the free volume of the E units in the facet node. By contrast, slip plane and Shockley shuffling involve shifts of entire atomic columns not only in the grain boundary itself, but also in the neighboring bulk crystals, requiring a larger degree of coordinated motion. Shifts involving atomic columns are restricted by the characteristics of local crystallographic planes, while atomic column dissociation releases the disordered nodes from those restrictions. Disordered shuffling can thus explain in large part why the Al boundaries generally migrate isotropically and offers a potential means of explaining the differing magnitudes of anisotropy observed in the Cu and Ni boundaries of Figure 5.

*4. Disordered cluster activity at facet nodes*

In the preceding sections, it was established that directionally-anisotropic mobility in Cu and Ni is a result of having a directionally-dependent shuffling mode (slip plane shuffling) that allows facet nodes to bypass Shockley partial contraction when migrating in one direction. This also allows Type A migrating facet nodes to shift more quickly between different shuffling modes, which includes disordered shuffling. Since Type B nodes are pinned by emitted Shockley partials for longer times, they are likely to have comparatively fewer E unit dissociation events and thus fewer occurrences of disordered shuffling per node. There is also a noted decrease in anisotropy and lag times with increasing temperature in the Cu and Ni boundaries (Figures 2 and 3), which we hypothesize is correlated with disordered shuffling becoming the dominant mode for both Type A and B motion at the highest homologous temperatures. Based on this reasoning, we propose that tracking the frequency of E unit dissociation and formation of disordered clusters occurring during migration can provide a means of quantifying how shifts in shuffling mode modulate directionally-anisotropic mobility in the Cu and Ni bicrystals.

To quantify the amount of disordered shuffling occurring per facet node during migration, an algorithm that tracks facet node dissociation events was developed. A representative example of the process for a Ni-1, 35.3° boundary during Type A motion is shown in Figure 8. First, the potential energy distribution of the grain boundary atoms is collected (Figure 8(a)). It was observed that, for each material, there is a particular atomic column within E units that has high potential energy but remains completely ordered. This atomic column for Ni-1 is shown in Figure 8(b) colored in blue, with an energy of approximately -4.58 eV (shown with a blue arrow in Figure 8(a)). Atoms with potential energies above this value tend to be involved in atomic column dissociation, shown in the grey shaded region in Figure 8(a). An example of a node with dissociation and selected high-energy atoms (black) is shown in Figure 8(c). In order to confine tracking only to completely disassociated atomic columns, atoms that are +0.1 eV higher than the intact atomic column are selected, which generally excluded partially-dissociated columns and shifts the focus to single high-energy/disordered atoms in the long tail of the potential energy distribution. Because dissociation events involve groups of atoms, the high-energy atoms may be spatially correlated and lead to an over- or under-counting of facet node clustering. To account for this, the Cluster Analysis algorithm in OVITO [46] is applied with a $1.1a_0$ cutoff to consolidate the spatial information. An example snapshot is shown in Figure 8(d), which identifies three separate clusters at two different nodes. From this analysis one may gain measure of the number of dissociation events occurring per timestep.

Results from this analysis for three example boundaries are shown in the plots of Figure 9(a-c), which each show the count of disordered clusters per node as a function of homologous temperature. As in earlier figures, the red data points show the counts for Type A motion and the blue curves those of Type B. The insets show each boundary's anisotropy ratio plots from Figure 2, with the data for the relevant inclination angle outlined in blue and highlighted. From Figures 9(a) and (b), it is clear that Type A migration has a consistently higher cluster count than Type B for all homologous temperatures. This data supports our hypothesis that there are directionally-dependent differences in the occurrence of E unit dissociation events, implying that disordered shuffling is occurring at a higher frequency in Type A migration. For both curves, increasing temperature leads to an increase in the counted clusters per node, which is also consistent with an increased rate of disordered shuffling for both motion directions. As mentioned earlier, Shockley partial contraction is a thermally-activated process and there is also a corresponding decrease in the Type B lag times shown in Figure 3. Most notably, there is a far more dramatic increase in cluster counts for Type B migration than Type A, which leads to a decrease in the difference between the cluster counts for the two types of motion (black arrow in Figure 9(a)). The decrease in this difference corresponds to large reductions in the anisotropy ratios for both the Cu-1 and Ni-1 boundaries. In contrast, the Al-1 boundary in Figure 9(c) has little to no difference in its Type A and B disordered cluster counts and, therefore, little to no anisotropy. These trends suggest that the cluster count per node may provide a means of connecting the anisotropy ratio directly to changes in boundary structure, namely the relative frequencies of Type A and Type B disordered shuffling.

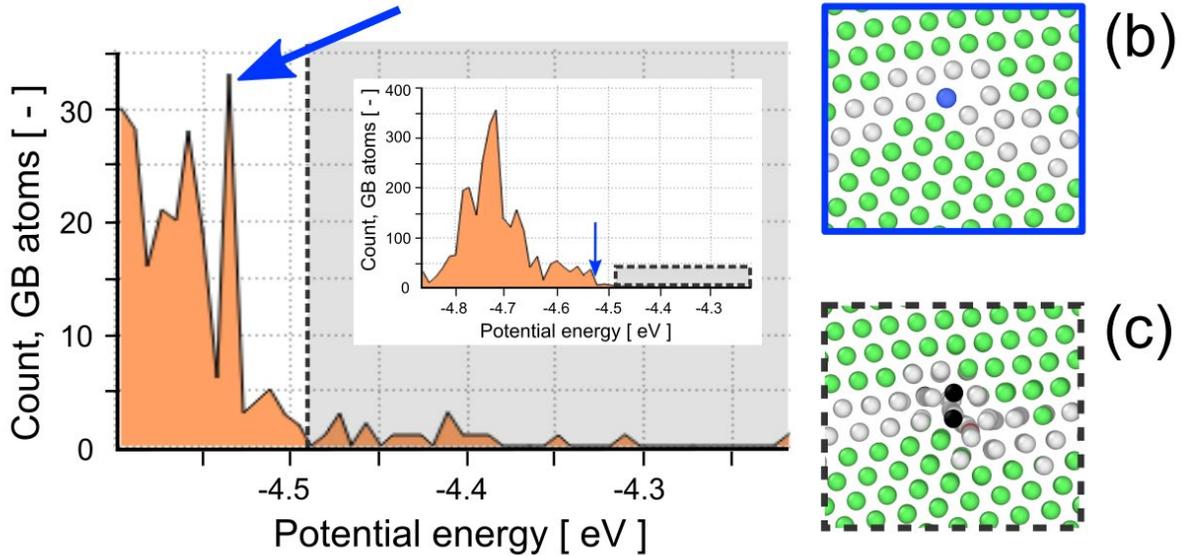

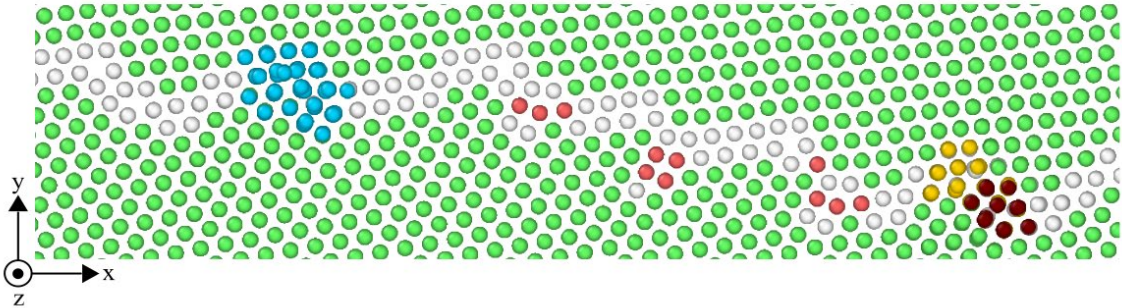

**Figure 8.** Overview of the disordered cluster identification algorithm used to track trends in disordered shuffling through all boundaries. (a) Example histogram showing the distribution of potential energies of a migrating Ni-1 boundary. The spike indicated by the blue arrow corresponds to the highest-energy fully intact atomic column in E units, shown in (b) as a blue atom. Atoms with higher potential energies above this spike tend to be involved in disordered shuffling events, which form clusters at facet nodes. (d) Final result of the cluster identification process for a particular boundary.

Figures 9(d-f) show the anisotropy plotted against the normalized disordered cluster count differences for each of the three homologous temperatures studied. In order to more clearly compare the difference in cluster counts across all 6 potentials and 4 inclination angles, the cluster counts were normalized against the maximum count for each combination of material, inclination angle, and temperature, yielding values between 0 and 1. Different colored data points indicate the different potentials, while inclination angles are shown using different symbol shapes. The two markers above the Y-axis limits in Figure 9(d) indicate the cluster size differences of the Ni-1 boundaries at 25.5° and 35.3° with anisotropy values of ~ 202 and ~50, respectively. The data indicate that there is a general, temperature-dependent positive correlation between the anisotropy

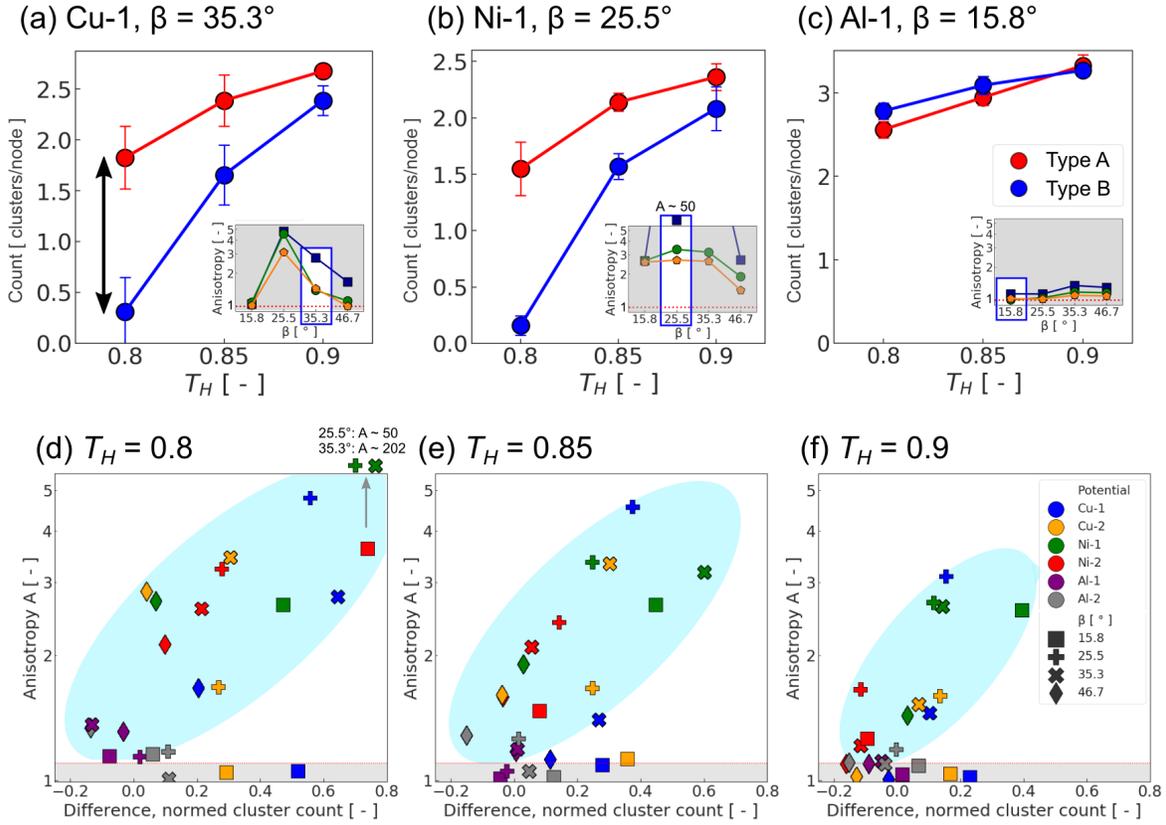

**Figure 9.** Normed disordered cluster counts for Type A (red) and Type B (blue) migration versus homologous temperature in example boundaries from (a) Cu, (b) Ni, and (c) Al. The average cluster counts exhibit a systematic difference between Type A and Type B (black arrow in (a) at $T_H = 0.8$) which decreases with increasing temperature. The inset shows the anisotropy mobility data for the chosen inclination angle angle (highlighted), showing a corresponding drop in anisotropy at $T_H = 0.85$ and 0.9. (d-f) Anisotropy versus the normed cluster count differences for each material and inclination angle. Blue regions are provided as a visual aid, enveloping boundaries with significant anisotropy values. The grey region in the bottom contains boundaries with little to no anisotropy (A < 1.1).

ratio and the Type A/B cluster count differences. This relationship is marked by a light blue oval in each plot, which we note begins above the anisotropy threshold of 1.1 (grey box). As temperature is increased from Figure 9(d) through (f), the light blue oval maintains its orientation, but becomes smaller and shorter. This also demonstrates that the cluster count differences are decreasing as the magnitude of anisotropy decreases. We thus conclude that temperature has the overall effect of increasing disordered shuffling at facet nodes, resulting in the corresponding reduction in anisotropy magnitudes.

There are still several trends in directionally-anisotropic mobility that have yet to be addressed. The weakly-anisotropic nature seen in some of the Al boundaries suggests that there may be a geometric component to atomic column dissociation in E units that is not yet accounted for, one that is especially sensitive to Type A migration. Potential evidence for this can be found in the work of Bowers et al. [45] on E units in an incommensurate boundary. These authors described Shockley contraction events as initiating with an agitation in the close-packed packed planes of one or two very specific columns. Though the crystallography of their incommensurate boundary is different, something similar could be occurring in the E units due to interactions between the {111}/{001} planes of neighboring IBP facets, especially given the changes in IBP facet length with changing inclination angles. Also left to explore in future work is the high anisotropy values seen in the Ni-1 boundary at $\beta = 25.5°$ and $\beta = 35.3°$, as well as the differences observed between the Ni and Cu boundaries at $\beta = 15.8°$. Those differences too could possibly arise from subtle differences between materials in the structure and dissociation behavior of E units. Lastly, the primary result of this work supports recent findings by Chen et al. [47] indicating that grain boundary mobility is most naturally expressed as a tensor quantity instead of a scalar. For these boundaries specifically, a mobility tensor would highlight the directional nature of migration velocity and could also provide unique insight into their unusual migration behavior through study of the individual tensor components.

## D. Summary and Conclusions

The migration behavior in a large set of faceted $\Sigma 11$ <110> tilt grain boundaries in three different face-centered cubic metals was investigated using a series of molecular dynamics simulations probing different potentials, temperatures, and inclination angles. From these results, the following conclusions can be made:

- Directionally-anisotropic mobility is discovered in a variety of different faceted asymmetric Σ11 <110> tilt boundaries in Cu and Ni. Boundaries with this anisotropy also exhibit a lag time before beginning motion, with some Ni boundaries barely moving within 1 ns. Boundaries were found to move up to 5 times faster in one direction than the other in many cases, with a few selected outliers showing high anisotropy ratios.

- Across 6 different materials/potentials, the same two structural units could be consistently identified: C units and E units. The atomic columns in E units can dissociate, forming clusters of disordered atoms at facet nodes. Due to the lower stacking fault energy in Cu and Ni boundaries, E units can also emit Shockley partials, which form E units once more when contracted.

- A low stacking fault energy material is necessary for directionally-anisotropic mobility to occur in these boundaries. In this case, the subjective classification for a "low" stacking fault energy means that a boundary must be able to emit grain boundary stacking faults, which excludes Al. Beyond this cutoff, the relationship between the magnitudes of stacking fault energy and anisotropy is unclear.

- Migration of the faceted boundaries is accomplished through transformation events at facet nodes and IBP facets, named *shuffling modes*. Three major shuffling modes were identified. *Shockley shuffling* is limited to materials that can emit and contract Shockley partials. *Slip plane shuffling* is a special IBP facet migration mode that only occurs during Type A motion due to the orientation of C units, which directly reflect the boundary's inclination angle $\beta$.

*Disordered shuffling* is common to all boundaries, and the only mode possible for Al, and involves the dissociation of E units into a mobile cluster of atoms.

- Disordered shuffling is the only mode which is, in general, directionally-isotropic with respect to mobility. The atoms involved in cluster motion have three degrees of translational motion and their displacements are localized to free volume within facet nodes. In contrast, Shockley shuffling and slip plane shuffling are both dependent on coordinated shifts of coherent atomic columns and thus necessarily involve local planes, which constrain motion in specific ways.
- In Cu and Ni, the rate-limiting mechanism of migration is Shockley partial contraction at facet nodes. If contraction can occur at a higher rate, the rate of node migration is increased. Factors that can increase contraction include increased temperatures (which increases the rate of isotropic disordered shuffling) and slip plane shuffling (which creates more favorable structural conditions for Shockley contraction). Directionally-anisotropic mobility in Cu and Ni can be primarily attributed to the increased node migration rate made possible in Type A motion by the slip plane shuffling mode.

The property of grain boundary mobility is commonly understood through the lens of bicrystallography (the five macroscopic degrees of freedom), which has proven to be an incredibly useful framework to date. However, the divergent mobilities seen in these faceted $\Sigma 11$ boundaries show that this description is likely not yet sufficient. Though directionally-anisotropic mobility is a phenomenon made possible by the crystallographic properties of the $\Sigma 11$ orientation, it is the complex interactions between those properties and the microscopic boundary structure, determined by energetic phenomena such as temperature and stacking fault energy, that ultimately give rise to it. There may be a wide range of yet undiscovered mobility phenomena that similarly arise from unique faceted boundary morphology.

## Acknowledgements

The authors would like to express their gratitude to Dr. Shawn P. Coleman and the U.S. Army Research Laboratory for providing the ECO force code. This work supported by the U.S. Department of Energy, Office of Basic Energy Sciences, Materials Science and Engineering Division under Award No. DE-SC0014232.